\documentclass[11pt]{article}

\usepackage[british]{babel}
\usepackage[T1]{fontenc}
\usepackage{times}
\usepackage{mathptmx}
\usepackage{amssymb}
\usepackage{amsmath}
\usepackage{amsthm}
\usepackage{exscale}

\numberwithin{equation}{section}

\makeatletter
\renewcommand{\@endtheorem}{\endtrivlist}
\makeatother

\theoremstyle{definition}
\newtheorem{Def}{Definition}[section]

\theoremstyle{plain}
\newtheorem{The}[Def]{Theorem}
\newtheorem{Pro}[Def]{Proposition}
\newtheorem{Cor}[Def]{Corollary}

\theoremstyle{remark}

\DeclareMathAlphabet{\mathib}{T1}{ptm}{b}{it}
\DeclareMathAlphabet{\mathscr}{U}{rsfs}{m}{n}

\DeclareMathOperator{\rank}{rank}
\DeclareMathOperator{\linhull}{span}
\DeclareMathOperator{\trace}{Tr \,}
\DeclareMathOperator{\id}{id}

\newcommand{\set}[1]{\{ #1 \}}
\newcommand{\bset}[1]{\bigl\{ #1 \bigr\}}

\newcommand{\abs}[1]{\lvert #1 \rvert}

\newcommand{\norm}[1]{\lVert #1 \rVert}
\newcommand{\bnorm}[1]{\bigl\lVert #1 \bigr\rVert}
\newcommand{\bgnorm}[1]{\biggl\lVert #1 \biggr\rVert}

\newcommand{\pnorm}[1]{\lVert #1 \rVert_p}

\newcommand{\nucl}[1]{\lVert #1 \rVert_1}

\newcommand{\bra}[1]{\langle #1 \vert}

\newcommand{\ket}[1]{\vert #1 \rangle}

\newcommand{\scp}[2]{\langle #1 \vert #2 \rangle}

\newcommand{\eg}{\textit{e.\,g.}}
\newcommand{\etc}{\textit{etc.}}
\newcommand{\ie}{\textit{i.\,e.}}

\newcommand{\Cbb}{\mathbb{C}}
\newcommand{\Kbb}{\mathbb{K}}
\newcommand{\Nbb}{\mathbb{N}}

\newcommand{\Efrak}{\mathfrak{E}}
\newcommand{\Ffrak}{\mathfrak{F}}

\newcommand{\Hscr}{\mathscr{H}}

\newcommand{\Oscr}{\mathscr{O}}

\newcommand{\BH}{\mathscr{B} ( \mathscr{H} )}
\newcommand{\BHone}{\mathscr{B} ( \mathscr{H}_1 )}
\newcommand{\BHtwo}{\mathscr{B} ( \mathscr{H}_2 )}

\newcommand{\AO}{\mathfrak{A} ( \mathscr{O} )}

\newcounter{defitem}

\newenvironment{deflist}{\begin{list}{(\alph{defitem})}%
  {\usecounter{defitem} \setlength{\topsep}{0ex}%
   \setlength{\parsep}{0.2ex} \setlength{\itemsep}{0.4ex}%
   \setlength{\leftmargin}{0em} \setlength{\itemindent}{0.5em}%
   }}{\end{list}}

\newcounter{proofitem}

\newenvironment{prooflist}{\begin{list}{(\roman{proofitem})}%
  {\usecounter{proofitem} \setlength{\topsep}{0ex}%
   \setlength{\labelwidth}{1em} \setlength{\labelsep}{0.5em}%
   \setlength{\parsep}{0.2ex} \setlength{\itemsep}{0.4ex}%
   \setlength{\leftmargin}{0em} \setlength{\itemindent}{1.5em}%
   \setlength{\listparindent}{1em}}}{\qed \end{list}}

\newcounter{romanitem}

\newenvironment{romanlist}{\begin{list}{\roman{romanitem})}%
  {\usecounter{romanitem} \setlength{\topsep}{1.5ex}%
   \setlength{\labelwidth}{1em} \setlength{\labelsep}{0.5em}%
   \setlength{\parsep}{0.2ex} \setlength{\itemsep}{0.4ex}%
   \setlength{\leftmargin}{0em} \setlength{\itemindent}{1.5em}%
   \setlength{\listparindent}{1em}}}{\end{list}}

\newcounter{alphitem}

\newenvironment{alphlist}{\begin{list}{\alph{alphitem})}%
  {\usecounter{alphitem} \setlength{\topsep}{1.5ex}%
   \setlength{\labelwidth}{1em} \setlength{\labelsep}{0.5em}%
   \setlength{\parsep}{0.2ex} \setlength{\itemsep}{0.4ex}%
   \setlength{\leftmargin}{0em} \setlength{\itemindent}{1.5em}%
   \setlength{\listparindent}{1em}}}{\end{list}}

\newcounter{arabicitem}

\newenvironment{arabiclist}{\begin{list}{(\arabic{arabicitem})}%
  {\usecounter{arabicitem} \setlength{\topsep}{0ex}%
   \setlength{\labelwidth}{1em} \setlength{\labelsep}{0.5em}%
   \setlength{\parsep}{0ex} \setlength{\itemsep}{0ex}%
   \setlength{\leftmargin}{1em} \setlength{\itemindent}{0.5em}%
   \setlength{\listparindent}{1em}}}{\end{list}}

\begin{document}

\begin{center}
  {\huge $p$-Nuclearity in a New Perspective}\\[1cm]
  {\large Christopher J. Fewster\\
    Department of Mathematics, University of York\\
    Heslington, York, YO10 5DD, United Kingdom\\
    email: cjf3@york.ac.uk\\[5mm]
    Izumi Ojima\\
    Research Institute for Mathematical Sciences, Kyoto University\\
    Kyoto 606-8502, Japan\\
    email: ojima@kurims.kyoto-u.ac.jp\\[5mm]
    Martin Porrmann\\
    II. Institut f\"ur Theoretische Physik, Universit\"at Hamburg\\
    Luruper Chaussee 149, 22761 Hamburg, Germany\\
    email: martin.porrmann@desy.de}
\end{center}

\begin{abstract}
  In this paper we try to settle some confused points concerning the
  use of the notion of $p$-nuclearity in the mathematical and physical
  literature, pointing out that the nuclearity index in the
  physicists' sense vanishes for any $p > 1$. Our discussion of these
  issues suggests a new perspective, in terms of $\epsilon$-entropy
  and operator spaces, which might permit connections to be drawn
  between phase space criteria and quantum energy inequalities.
\end{abstract}

\section{Introduction}

The Araki--Haag--Kastler programme of algebraic quantum field theory
\cite{haag:1996} seeks to describe theories in terms of algebras $\AO$
of observables associated with open bound\-ed regions $\Oscr$ in
spacetime, with particular regard to their net structure encoded by
the map $\Oscr \mapsto \AO$. The structural analysis of general
quantum field theories within this framework proceeds from a small
number of axioms relating to Poincar\'e covariance and causality.
However, these axioms by themselves do not guarantee reasonable
physical behaviour, such as the existence of thermodynamical
equilibrium states or particle-like excitations. For these properties,
it turns out that one must impose additional restrictions on the phase
space volume available to the theory, according to some suitable
notion of size. Criteria of this type were first introduced by Haag
and Swieca \cite{haag/swieca:1965} in terms of a compactness
condition, and have since proved their worth in various contexts. In
particular, the nuclearity criterion of Buchholz and Wichmann
\cite{buchholz/wichmann:1986} was applied to the analysis of
thermodynamic properties, \eg\ in \cite{buchholz/junglas:1989}, and to
modular theory \cite{buchholz/dantoni/longo:1990}.

A typical (but by no means the only) setting for these criteria is the
following. Two Banach spaces $\Efrak$ and $\Ffrak$ are identified,
along with a class of continuous maps $\Theta_{\beta , \Oscr}: \Efrak
\rightarrow \Ffrak$ which are associated with energetically damped
local excitations of the vacuum, localised in $\Oscr$ and with the
damping parametrised by the inverse temperature $\beta$. The phase
space requirement is then encoded by demanding that these maps be
`approximately finite rank': more precisely, that they belong to some
class of operators containing the finite rank maps, and contained
within the class of compact maps from $\Efrak$ to $\Ffrak$. A
numerical index, $\nu$, is defined on this class of operators, and the
asymptotic behaviour of $\nu ( \Theta_{\beta , \Oscr} )$ as $\beta \to
0^+$ may also be constrained as part of the nuclearity criterion. The
main purpose of this paper is to draw attention to a serious
shortcoming of one such index, namely the so-called $p$-nuclearity
index. Indeed, we will show that this index vanishes identically for
$p > 1$!

The starting point for our discussion is the investigation of
\cite{buchholz/dantoni/longo:1990}, in which mappings of type $l^p$
are considered\footnote{See Sec. 2 for the definition of this class}
and, using the result \cite[8.4.2~Proposition]{pietsch:1972}, seen to
be nuclear for $0 < p \leqslant 1$. In fact, there exists a
decomposition of a mapping $\Theta : \Efrak \rightarrow \Ffrak$ of
type $l^p$ ($\Efrak$ and $\Ffrak$ normed vector spaces) in terms of
sequences of vectors $\bset{\varphi_k : k \in \Nbb} \subset \Ffrak$
and of continuous linear functionals $\bset{\ell_k : k \in \Nbb}
\subset \Efrak^*$ such that
\begin{subequations}
  \begin{eqnarray}
    \Theta ( x ) = \sum_{k = 1}^\infty \ell_k ( x ) \varphi_k
    \text{,} \quad x \in \Efrak \text{,}\\
    \text{and} \quad \sum_{k = 1}^\infty \norm{\ell_k}^p
    \norm{\varphi_k}^p < \infty \text{.}
  \end{eqnarray} 
  The corresponding number
  \begin{equation}
    \pnorm{\Theta} \doteq
    \inf \left( \sum_{k = 1}^\infty \bnorm{\ell_k}^p
    \bnorm{\varphi_k}^p \right)^{\frac{1}{p}} \text{,}
  \end{equation}
\end{subequations}
defines a quasi-norm on the set of these mappings which is called the
$p$-norm, where the infimum extends over all possible decompositions
of $\Theta$. Now the map $\Theta$ is said to be nuclear if
$\|\Theta\|_1$ is finite, and, since $\|\Theta\|_p \geqslant
\|\Theta\|_1$ for $0 < p \leqslant 1$, the $l^p$ maps are indeed
nuclear for $p$ in this interval.

For $p \leqslant 1$ the above definition is sound and indeed has an
intuitive interpretation in algebraic QFT. Moreover, its use has led
to deep insights as already mentioned. However, later authors have
often adopted the above notions without any restriction on $p$, under
the label of \textit{{p}-nuclearity}. The first occurrence may be in
\cite[Section~2]{buchholz/porrmann:1990}, further examples of this
usage may be found in
\cite{buchholz/dantoni:1995,buchholz:1996,mohrdieck:2002,%
  dantoni/hollands:2004}. But it ought to be emphasised that the
physically relevant results published in these papers depend on the
case $p \leqslant 1$ and therefore remain valid, despite the problems
with $p > 1$ we will point out. The explicit definition of this
version, following Buchholz, D'Antoni, and Longo in
\cite{buchholz/dantoni/longo:1990} (who, however, did \textit{not} use
the term $p$-nuclearity in this context) can be formulated as follows.
\begin{Def}
  \label{Def:p-nuclearity-phys}
  Let $\Efrak$ and $\Ffrak$ be normed vector spaces. An operator
  $\Theta : \Efrak \to \Ffrak$ is called
  \textit{\bfseries$\mathib{p}$-nuclear}, $p > 0$, if there exist
  sequences of vectors $\bset{\varphi_k : k \in \Nbb} \subset \Ffrak$
  and of continuous linear functionals $\bset{\ell_k : k \in \Nbb}
  \subset \Efrak^*$ such that
  \begin{subequations}
    \begin{eqnarray}
      \label{eq:p-nucl-1}
      \Theta ( x ) = \sum_{k = 1}^\infty \ell_k ( x ) \varphi_k
      \text{,} \quad x \in \Efrak \text{,}\\
      \label{eq:p-nucl-2}
      \text{and} \quad \sum_{k = 1}^\infty \norm{\ell_k}^p
      \norm{\varphi_k}^p < \infty \text{.}
    \end{eqnarray}
    For later reference we call a combination of functionals and
    vectors satisfying the relations \eqref{eq:p-nucl-1} and
    \eqref{eq:p-nucl-2} a \textit{\bfseries$\mathib{p}$-nuclear
      decomposition of $\Theta$}.  The $p$-nuclearity index of the
    operator $\Theta$ is defined by
    \begin{equation}
      \label{eq:p-nucl-index}
      \pnorm{\Theta} \doteq
      \inf_{\substack{\text{$p$-nuclear}\\\text{decompositions}}} 
      \left( \sum_{k = 1}^\infty \bnorm{\ell_k}^p \bnorm{\varphi_k}^p
      \right)^{\frac{1}{p}} \text{.}
    \end{equation}
  \end{subequations}
\end{Def}

Now it is true that mappings of type $l^p$ are $p$-nuclear for
arbitrary $p > 0$, as we will show in the next section by reworking
Pietsch's argument. But two important caveats should be borne in mind.
First, there is a notion of $p$-nuclearity for $p \geqslant 1$ in the
mathematical literature which differs from that given above (to which
we shall refer as the physicists' definition) except in the case
$p = 1$. Second, the $p$-nuclearity index of \eqref{eq:p-nucl-index}
can easily be seen to vanish for any $p$-nuclear operator if $p > 1$.
Furthermore, if $\Efrak$ is a Banach space with a Schauder basis then
every bounded operator from $\Efrak$ to $\Ffrak$ is $p$-nuclear for
all $p > 1$ (with necessarily vanishing $p$-nuclearity index). To the
best of our knowledge this has not been pointed out before.

The object of this letter is to clarify the above issues and also to
indicate a possible remedy for the problem just mentioned, along with
alternative directions for research. Our investigation was inspired by
the wish to use phase space criteria like nuclearity to establish
\textit{quantum energy inequalities} (QEIs). These are state-independent
lower bounds on weighted averages of the stress-energy tensor, which
have been established for various free field theories and
two-dimensional conformal field theory (see
\cite{fewster/hollands:2004} and references therein).  Now QEIs are a
manifestation of the uncertainty principle, and therefore intimately
related to phase space properties of the theory (see
\cite{eveson/fewster/verch:2005} for quantum mechanical examples of
this connection). It is therefore natural to enquire whether there is
a more formal connection between QEIs and nuclearity criteria. Some
progress on this question has already been made, in the context of
generalised free fields with discrete mass spectrum, and will be
reported in full elsewhere. It turns out that the existence of QEIs
with reasonable scaling behaviour are equivalent to growth conditions
on the mass spectrum which are sufficient for nuclearity to hold with
the correct asymptotic behaviour of the nuclearity index. In order to
establish full equivalence between QEIs and nuclearity, it is
necessary to obtain lower bounds on the nuclearity index. [Upper
bounds are of course provided by any decomposition entering the
definition of the $p$-norm in \eqref{eq:p-nucl-index}.] As a first
step in this direction we indicate an \textit{exact} expression for the
$2$-nuclearity index of a $2$-nuclear operator acting between Hilbert
spaces, using a modified notion of $2$-nuclearity. Another possible
route from phase space criteria to quantum inequalities could be the
use of the notion of $\epsilon$-entropy ($\epsilon$-content) in the
context of operator spaces. Not only upper but also lower bounds on
the $\epsilon$-entropy can be defined in the limit of small $\epsilon$
\cite{akashi:1990}. It is hoped to return to these issues elsewhere.

\section{Decompositions of Mappings of Type
  $\mathib{l}^{\thinspace\mathib{p}}$ for Arbitrary $\mathib{p}$ and
  Related Problems}
  \label{sec:l-p-nucl}

We begin with the formal definition of mappings of type $l^p$.
\begin{Def}[Pietsch {\cite[8.1.1]{pietsch:1972}}]
  Let $\Efrak$ and $\Ffrak$ be normed vector spaces. For an arbitrary
  continuous operator $\Theta : \Efrak \to \Ffrak$ we define the
  \textit{\bfseries$\mathib{k}$'th approximation number} $\alpha_k (
  \Theta )$, $k \in \Nbb_0$, through
  \begin{equation}
    \label{eq:approx-number}
    \alpha_k ( \Theta ) \doteq \inf \bset{\norm{\Theta - \Theta_k} :
      \text{$\Theta_k$ an operator of at most finite rank $k$}}
      \text{.}  
  \end{equation}
\end{Def}
These approximation numbers can now be used to define certain
subspaces $l^p ( \Efrak , \Ffrak )$ of continuous operators.
\begin{Def}[Pietsch {\cite[8.2.1]{pietsch:1972}}]
  \label{Def:type-l-p}
  A continuous operator $\Theta : \Efrak \to \Ffrak$ is said to be a
  \textit{\bfseries mapping of type
    $\mathib{l}^{\thinspace\mathib{p}}$} (called \textit{\bfseries
    $\mathib{p}$-approximable operator} in \cite[19.8]{jarchow:1981}),
  $0 < p < \infty$, if summation of the $p$'th power of all
  approximation numbers yields a finite result:
  \begin{equation}
    \label{eq:l-p-cond}
    \sum_{k = 0}^\infty \alpha_k ( \Theta )^p < \infty \text{.}
  \end{equation}
  Furthermore one defines the real number
  \begin{equation}
    \label{eq:rho-p}
    \rho_p ( \Theta ) \doteq \left( \sum_{k = 0}^\infty \alpha_k (
      \Theta )^p \right)^{\frac{1}{p}} \text{.}
  \end{equation}
\end{Def}

We now generalise the result of Pietsch
\cite[8.4.2~Proposition]{pietsch:1972} mentioned above to all $p > 0$,
re-writing his proof in a notation more familiar to physicists.
\begin{Pro}
  \label{Pro:l-p-nuc}
  For all $p > 0$ each mapping $\Theta \in l^p ( \Efrak , \Ffrak )$
  can be represented as
  \begin{equation}
    \label{eq:l-p-nuc}
    \Theta ( x ) = \sum_{k = 1}^\infty \lambda_k \ell_k ( x ) \varphi_k
    \text{,} \quad x \in \Efrak \text{,}
  \end{equation}
  with normalised sequences of vectors $\bset{\varphi_k : k \in \Nbb}
  \subset \Ffrak$ and of continuous linear functionals $\bset{\ell_k :
  k \in \Nbb} \subset \Efrak^*$ as well as of numbers $0 < \lambda_k
  \leqslant \norm{\Theta}$ such that
  \begin{equation}
    \label{eq:l-p-nuc-estimate}
    \left( \sum_{k = 0}^\infty \lambda_k^p \right)^{\frac{1}{p}}
    \leqslant 2^{2 + \frac{3}{p}} \rho_p ( \Theta ) \text{.}
  \end{equation}
\end{Pro}
\begin{proof}
  Consider approximations $\Theta_n$ of rank $2^n -2$ which satisfy
  $\norm{\Theta - \Theta_n} \leqslant 2 \alpha_{2^n -2} ( \Theta
  )$. (Note that $\Theta_1 = 0$.) Then $\Psi_n \doteq \Theta_{n + 1} -
  \Theta_n$ is an operator of rank $\leqslant 2^{n+2}$ with
  \begin{equation*}
    \norm{\Psi_n} \leqslant \norm{\Theta - \Theta_n} - \norm{\Theta -
    \Theta_{n + 1}} \leqslant 4 \alpha_{2^n -2} ( \Theta ) \text{,}
  \end{equation*}
  since the sequence of approximation numbers is monotone
  decreasing, so
  \begin{equation*}
    \rank ( \Psi_n ) \norm{\Psi_n}^p \leqslant 2^{n +
    2} 4^p \alpha_{2^n -2} ( \Theta )^p = 2^{2 p + n +2} \alpha_{2^n
    -2} ( \Theta )^p \text{.}
  \end{equation*}
  Now the monotone decrease of the approximation numbers permits us to
  use Cau\-chy's condensation trick to write
  \begin{equation*}
    \sum_{n = 1}^\infty 2^{n - 1} \alpha_{2^n -2} ( \Theta )^p
    \leqslant \sum_{n = 1}^\infty \sum_{r = 2^{n -1} - 1}^{2^n -2}
    \alpha_r ( \Theta )^p = \sum_{r = 0}^\infty \alpha_r ( \Theta )^p
    = \rho_p ( \Theta )^p \text{.}
  \end{equation*}
  Hence
  \begin{equation*}
    \label{eq:sum-lambda}
    \sum_{n = 1}^\infty \rank ( \Psi_n ) \norm{\Psi_n}^p \leqslant
    2^{2 p + 3} \sum_{n = 1}^\infty 2^{n - 1} \alpha_{2^n -2} ( \Theta
    )^p \leqslant 2^{2 p + 3} \rho_p ( \Theta )^p \text{.}
  \end{equation*}
  According to \cite[8.4.1, Lemma~2]{pietsch:1972} $\Psi_n$ as an
  operator of finite rank can be written as
  \begin{equation*}
    \label{eq:n-th-summand}
    \Psi_n ( x ) = \sum_{i = 1}^{\rank ( \Psi_n )} \lambda_i^{( n )}
    \ell_i^{( n )} ( x ) \varphi_i^{( n )} \text{,} \quad x \in \Efrak
    \text{,}
  \end{equation*}
  with $0 < \lambda_i^{( n )} \leqslant \norm{\Psi_n}$ and normalised
  functionals $\ell_i^{( n )}$ and vectors $\varphi_i^{( n )}$ in
  $\Efrak^*$ and $\Ffrak$, respectively. Moreover,
  \begin{equation}
    \label{eq:lambda-sum-estimate}
    \sum_{n = 1}^\infty \sum_{i = 1}^{\rank ( \Psi_n )} \left(
    \lambda_i^{( n )} \right)^p \leqslant \sum_{n = 1}^\infty \rank (
    \Psi_n ) \norm{\Psi_n}^p \leqslant 2^{2 p + 3} \rho_p ( \Theta )^p
    \text{.} 
  \end{equation}
  By definition,
  \begin{equation}
    \label{eq:Theta-rep}
    \Theta ( x ) = \lim_{m \rightarrow \infty} \Theta_m ( x ) = \lim_{m
    \rightarrow \infty} \sum_{n = 1}^m \Psi_n ( x ) = \sum_{n =
    1}^\infty \sum_{i = 1}^{\rank ( \Psi_n )} \lambda_i^{( n )}
    \ell_i^{( n )} ( x ) \varphi_i^{( n )} \text{,} \quad x \in \Efrak
    \text{,}
  \end{equation}
  which in connection with \eqref{eq:lambda-sum-estimate} establishes
  the Proposition, without any restriction on $p$.
\end{proof}

Since we are free to choose functionals and vectors that are not
normalised in the representation \eqref{eq:l-p-nuc} of the operator
$\Theta$, absorbing the coefficients $\lambda_k$ into one or both of
them, the above result shows that in this case the product of the
norms raised to the $p$'th power is summable with the bound given in
\eqref{eq:l-p-nuc-estimate}. Thus we have the following Corollary.
\begin{Cor}
  \label{Cor:p-nucl}
  Every operator $\Theta \in l^p ( \Efrak , \Ffrak )$ is $p$-nuclear
  in the sense of Definition~\ref{Def:p-nuclearity-phys} for arbitrary
  $p > 0$.
\end{Cor}

Despite this close relationship with mappings of type $l^p$, the
notion of $p$-nuclearity as defined in
Def.~\ref{Def:p-nuclearity-phys} is problematic in two ways. First of
all it is in conflict with the mathematicians' notion that is defined
for $1 \leqslant p \leqslant \infty$. For completeness we present the
formal definition following Jarchow's book \cite{jarchow:1981} and
adopt a notation hopefully better accessible for mathematical
physicists.
\begin{Def}
  \label{Def:p-nuclearity-math}
  Let $\Efrak$ and $\Ffrak$ be normed vector spaces. An operator
  $\Theta : \Efrak \to \Ffrak$ is called
  \textit{\bfseries$\mathib{p}$-nuclear}, $1 \leqslant p \leqslant
  \infty$, if there exist sequences of vectors $\bset{\varphi_k : k
  \in \Nbb} \subset \Ffrak$ and of continuous linear functionals
  $\bset{\ell_k : k \in \Nbb} \subset \Efrak^*$ such that
  \begin{equation}
    \label{eq:p-nucl-math}
    \Theta ( x ) = \sum_{k = 1}^\infty \ell_k ( x ) \varphi_k
    \text{,} \quad x \in \Efrak \text{,}
  \end{equation}
  and such that the sequences comply with the following additional
  assumptions. Again we refer to such a decomposition as a
  \textit{\bfseries$\mathib{p}$-nuclear decomposition of $\Theta$} in
  each case.
  \begin{deflist}
  \item\label{Defsub:p-nucl} For $1 < p < \infty$ there hold
    \begin{subequations}
      \begin{eqnarray}
        \label{eq:p-nucl-sequ-1}
        \sum_{k = 1}^\infty \norm{\ell_k}^p < \infty \quad
        \text{and}\\
        \label{eq:p-nucl-sequ-2}
        \sum_{k = 1}^\infty \abs{\scp{\alpha}{\varphi_k}}^{p^*} <
        \infty \text{,} \quad \alpha \in \Ffrak^* \text{.}
      \end{eqnarray}
      Here $p^*$ is the conjugate number to $p$, \ie, $p^* \doteq
      \frac{p}{p - 1}$ so that $p^{- 1} + {p^*}^{- 1} = 1$. The
      $p$-nuclearity index $\nu_p ( \Theta )$ is then defined by
      \begin{equation}
        \label{eq:p-nucl-index-math}
        \nu_p ( \Theta ) \doteq
        \inf_{\substack{\text{$p$-nuclear}\\\text{decompositions}}}
        \left( \left( \sum_{k = 1}^\infty \norm{\ell_k}^p
        \right)^{\frac{1}{p}} \cdot \sup_{\substack{\alpha \in
        \Ffrak^*\\\norm{\alpha} \leqslant 1}} \left( \sum_{k =
        1}^\infty \abs{\scp{\alpha}{\varphi_k}}^{p^*}
        \right)^{\frac{1}{p^*}} \right) \text{.}
      \end{equation}
    \end{subequations}
  \item\label{Defsub:1-nucl} For $p = 1$ ($1^* = \infty$) the
    additional conditions on the sequences are
    \begin{subequations}
      \begin{eqnarray}
        \label{eq:1-nucl-sequ-1}
        \sum_{k = 1}^\infty \norm{\ell_k} < \infty \quad \text{and}\\
        \label{eq:1-nucl-sequ-2}
        \sup_{k \in \Nbb} \; \norm{\varphi_k} < \infty \text{,}
      \end{eqnarray}
      with the $1$-nuclearity index $\nu_1 ( \Theta )$ defined by
      \begin{equation}
        \label{eq:1-nucl-index-math}
        \nu_1 ( \Theta ) \doteq
        \inf_{\substack{\text{$1$-nuclear}\\\text{decompositions}}}
        \left( \left( \sum_{k = 1}^\infty \norm{\ell_k} \right)
        \cdot \sup_{k \in \Nbb} \; \norm{\varphi_k} \right) \text{.}
      \end{equation}
    \end{subequations}
  \item\label{Defsub:inf-nucl} For $p = \infty$ ($\infty^* = 1$) the
    conditions on the sequences are
    \begin{subequations}
      \begin{eqnarray}
        \label{eq:inf-nucl-sequ-1}
        \lim_{k \to \infty} \norm{\ell_k} = 0 \quad \text{and}\\
        \label{eq:inf-nucl-sequ-2}
        \sum_{k = 1}^\infty \abs{\scp{\alpha}{\varphi_k}} < \infty
        \text{,} \quad \alpha \in \Ffrak^* \text{,}
      \end{eqnarray}
      with the $\infty$-nuclearity index $\nu_\infty ( \Theta )$
      defined as
      \begin{equation}
        \label{eq:inf-nucl-index-math}
        \nu_\infty ( \Theta ) \doteq
        \inf_{\substack{\text{$\infty$-nuclear}\\\text{decompositions}}}
        \left( \left( \sup_{k \in \Nbb} \; \norm{\ell_k} \right) \cdot
        \sup_{\substack{\alpha \in \Ffrak^*\\\norm{\alpha} \leqslant
        1}} \left( \sum_{k = 1}^\infty \abs{\scp{\alpha}{\varphi_k}}
        \right) \right) \text{.}
      \end{equation}
    \end{subequations}
  \end{deflist}
  Note that the definitions \eqref{eq:p-nucl-index-math} and
  \eqref{eq:inf-nucl-index-math} indeed yield finite values for the
  nuclearity index, since 
  \begin{equation*}
    \sup_{\substack{\alpha \in \Ffrak^*\\\norm{\alpha} \leqslant 1}}
    \left( \sum_{k = 1}^\infty \abs{\scp{\alpha}{\varphi_k}}^p \right)
    < \infty
  \end{equation*}
  for any $p > 0$ by a uniform boundedness argument
  (cf.~\cite[Section~16.5]{jarchow:1981} for the case $p = 1$).
\end{Def}

In the case $p = 1$ both notions of nuclearity coincide.
\begin{Pro}
  \label{Pro:1-nucl-index}
  The nuclearity index for $p = 1$ calculated according to
  \eqref{eq:p-nucl-index} and \eqref{eq:1-nucl-index-math} yields the
  same result.
\end{Pro}
\begin{proof}
  \begin{prooflist}
  \item Let $\bset{\varphi_k : k \in \Nbb} \subset \Ffrak$ and
    $\bset{\ell_k : k \in \Nbb} \subset \Efrak^*$ be sequences of
    vectors and of continuous linear functionals, respectively,
    complying with relations \eqref{eq:p-nucl-1} and
    \eqref{eq:p-nucl-2} for $p = 1$. A simple redefinition by
    $\ell_k^\prime \doteq \norm{\varphi_k} \; \ell_k$ and
    $\varphi_k^\prime \doteq \norm{\varphi_k}^{- 1} \varphi_k$ yields
    another $1$-nuclear decomposition of $\Theta$ in terms of unit
    vectors satisfying
    \begin{equation*}
      \sum_{k = 1}^\infty \norm{\ell_k} \norm{\varphi_k} = \sum_{k =
      1}^\infty \norm{\ell_k^\prime} \norm{\varphi_k^\prime} = \sum_{k
      = 1}^\infty \norm{\ell_k^\prime} \text{.} 
    \end{equation*}
    From this relation we not only infer that the nuclearity index
    $\nucl{\Theta}$ can be calculated by considering only such
    $1$-nuclear decompositions in terms of sequences of unit vectors.
    Furthermore, these special decompositions comply with the
    requirements stated in
    Definition~\ref{Def:p-nuclearity-math}\ref{Defsub:1-nucl} with
    $\sup_{k \in \Nbb} \; \norm{\varphi_k^\prime} = 1$ so that,
    according to \eqref{eq:1-nucl-index-math},
    \begin{equation*}
      \nu_1 ( \Theta ) \leqslant \sum_{k = 1}^\infty
      \norm{\ell_k^\prime} \text{.}
    \end{equation*}
    Thus we conclude that
    \begin{equation*}
      \nu_1 ( \Theta ) \leqslant \nucl{\Theta} \text{.}
    \end{equation*}
  \item Now, let $\bset{\varphi_k : k \in \Nbb} \subset \Ffrak$ and
    $\bset{\ell_k : k \in \Nbb} \subset \Efrak^*$ be sequences of
    vectors and of continuous linear functionals, respectively,
    complying with equation \eqref{eq:p-nucl-math} and the
    requirements \eqref{eq:1-nucl-sequ-1} and
    \eqref{eq:1-nucl-sequ-2}. Then,
    \begin{equation*}
      \sum_{k = 1}^\infty \bnorm{\ell_k} \bnorm{\varphi_k} \leqslant
      \left( \sum_{k = 1}^\infty \norm{\ell_k} \right) \cdot \sup_{k
      \in \Nbb} \; \norm{\varphi_k} < \infty \text{,}
    \end{equation*}
    so that these sequences conform to \eqref{eq:p-nucl-2} in
    Definition~\ref{Def:p-nuclearity-phys} for $p = 1$. Moreover, we
    conclude from this relation in connection with
    \eqref{eq:p-nucl-index} and \eqref{eq:1-nucl-index-math} that
    \begin{equation*}
      \nucl{\Theta} \leqslant \nu_1 ( \Theta ) \text{.}
    \end{equation*}
    \renewcommand{\qed}{}
  \end{prooflist}
  Combining both results we arrive at the desired statement
  \begin{equation*}
    \nucl{\Theta} = \nu_1 ( \Theta ) 
  \end{equation*}
  which is valid for any operator $\Theta$ complying with either
  Definition~\ref{Def:p-nuclearity-phys} for $p = 1$ or with
  Definition~\ref{Def:p-nuclearity-math}\ref{Defsub:1-nucl}.
\end{proof}

But the conflict with the mathematicians' concept is not the only
problem connected with Definition~\ref{Def:p-nuclearity-phys}. In fact
as it stands the corresponding $p$-nuclearity index is identically
zero for $p > 1$, due to the following reasoning. Since no further
restriction is imposed on the sequence of vectors $\bset{\varphi_k : k
  \in \Nbb} \subset \Ffrak$ appearing in \eqref{eq:p-nucl-1}, we are
free to replace every term in this sum by $m$ equal terms consisting
in the product of $\ell_k ( x )$ with $m^{- 1} \varphi_k$. The result
is another $p$-nuclear decomposition for $\Theta$.  But every term
$\norm{\ell_k}^p \norm{\varphi_k}^p$ in \eqref{eq:p-nucl-2} is now
replaced by $m$ identical terms $\norm{\ell_k}^p \norm{m^{- 1}
  \varphi_k}^p = m^{- p} \norm{\ell_k}^p \norm{\varphi_k}^p$. In this
way, the sum in \eqref{eq:p-nucl-2} is multiplied by the factor $m
\cdot m^{- p} = m^{- ( p - 1 )}$ yielding another upper bound for the
$p$-nuclearity index defined in \eqref{eq:p-nucl-index} which is the
original one times $m^{- \frac{p - 1}{p}}$. Since we are free to
choose an arbitrary large natural number $m$, we thus get arbitrarily
low upper bounds for the $p$-nuclearity index defined according to
\eqref{eq:p-nucl-index} in the parameter range $1 < p < \infty$. In
fact, $\pnorm{\Theta} = 0$, which gives no insight into the
geometrical structure of $\Theta ( \Efrak_1 )$ as was originally
hoped.  For the range $0 < p \leqslant 1$ no such problem with the
$p$-nuclearity index $\pnorm{~.~}$ arises; and on the bounds of the
index $\nu_p (~.~)$ according to \eqref{eq:p-nucl-index-math} in
Definition~\ref{Def:p-nuclearity-math}\ref{Defsub:p-nucl} the
artificial splitting of individual terms in a given $p$-nuclear
decomposition has no effect at all.

The above procedure can be used to produce an even more striking
result when the pre-image $\Efrak$ of the operator $\Theta$ is an
infinite-dimensional Banach space with Schauder basis (in
contradistinction to a Hamel basis), \ie, when every element of
$\Efrak$ has a decomposition $x = \sum_{k = 1}^\infty \alpha_k x_k$ in
terms of a fixed sequence $\set{x_k}_{k \in \Nbb} \subseteq \Efrak$
with a \textit{unique} set of coefficients $\set{\alpha_k}_{k \in
  \Nbb}$ \cite{schauder:1927a}.
\begin{Pro}
  \label{Pro:schauder}
  Every bounded operator $\Theta$ mapping the Banach space $\Efrak$
  with Schauder basis $\set{x_k}_{k \in \Nbb}$ into the normed space
  $\Ffrak$ is $p$-nuclear for every $p > 1$ in the sense of
  Definition~\ref{Def:p-nuclearity-phys}.
\end{Pro}
\begin{proof}
  Let $\set{\varphi_k}_{k \in \Nbb}$ denote the sequence of images of
  elements of the Schauder basis in $\Efrak$ under $\Theta$:
  $\varphi_k \doteq \Theta ( x_k )$, $k \in \Nbb$. Let furthermore
  $\set{\ell_k}_{k \in \Nbb} \subseteq \Efrak^*$ denote the sequence
  of associated (continuous) coefficient functionals defined via
  $\ell_k ( x ) \doteq \alpha_k$, $k \in \Nbb$, $x \in \Efrak$, where
  $x = \sum_{k = 1}^\infty \alpha_k x_k$ \cite[Definition 3.1 and
  Theorem 3.1]{singer:1970}. By linearity and continuity of $\Theta$,
  we then get the following decomposition of $\Theta ( x )$:
  \begin{equation}
    \label{eq:schauder-rep}
    \Theta ( x ) = \sum_{k = 1}^\infty \ell_k ( x ) \Theta ( x_k ) =
    \sum_{k = 1}^\infty \ell_k ( x ) \varphi_k \text{,} \quad x \in
    \Efrak \text{.}
  \end{equation}
  While being a decomposition of the desired kind \eqref{eq:p-nucl-1},
  the validity of relation \eqref{eq:p-nucl-2} is not guaranteed. But
  this can be achieved by use of the same sort of dilution argument
  that was already applied in the paragraph preceding
  Proposition~\ref{Pro:schauder}. Since $p$ is supposed to be greater
  than $1$ we can exhibit for every $k \in \Nbb$ a natural number
  $m_k$ such that
  \begin{equation}
    \label{eq:schauder-estimate}
    m_k^{1-p} \norm{\ell_k}^p \norm{\phi_k}^p < \frac{1}{k^2} \text{.}
  \end{equation}
  Replacing the $k$'th term in \eqref{eq:schauder-rep} by $m_k$
  identical copies $m_k^{- 1} \ell_k ( x ) \varphi_k$ we get another
  decomposition of $\Theta ( x )$ compliant with \eqref{eq:p-nucl-1};
  but now in addition \eqref{eq:p-nucl-2} is satisfied since,
  according to \eqref{eq:schauder-estimate},
  \begin{equation}
    \label{eq:schauder-rep-ext}
    \sum_{k = 1}^\infty m_k \norm{m_k^{- 1} \ell_k}^p
    \norm{\varphi_k}^p = \sum_{k = 1}^\infty m_k^{1-p} \norm{\ell_k}^p
    \norm{\phi_k}^p < \sum_{k = 1}^\infty \frac{1}{k^2} =
    \frac{\pi^2}{6} < \infty \text{.}
  \end{equation}
  Thus, $\Theta$ turns out to be $p$-nuclear for every $p > 1$. 
\end{proof}
Again we perceive that the notion of $p$-nuclearity as formulated in
Definition~\ref{Def:p-nuclearity-phys} loses its discriminatory power
for $p > 1$.

A natural first reaction to these problems is to insist that the
sequences of vectors occurring in $p$-nuclear decompostions should be
linearly independent. However, this does not provide a remedy, as we
now show. Suppose that the span of the vectors $\varphi_j$ appearing
in the nuclear decomposition of $\Theta$ has infinite codimension in
$\Ffrak$ and choose countably many sequences of countably many vectors
$\xi_{r,s}$, $r$, $s \in \Nbb$, which are linearly independent of each
other and the $\varphi_j$. The aim is to show that we can modify the
decomposition of $\Theta$ by using only the $\xi_{1,s}$'s, so that the
upper bound on $\pnorm{\Theta}$ is reduced by a factor strictly less
than one (independent of the $\xi$'s). By repeating this, using the
$\xi_{2,s}$'s \etc, the upper bound becomes arbitrarily close to zero.
To do this, choose $\alpha$ so that $\frac{1}{2} < \alpha < 2^{-
  \frac{1}{p}}$ which is possible for $p > 1$.  We may assume without
loss of generality that the $\xi_{1,s}$'s have been normalised so that
\begin{equation*}
  \Phi_{j , \pm} \doteq \frac{1}{2} \varphi_j \pm \xi_{1,j}
\end{equation*}
have norms $\norm{\Phi_{j , \pm}} \leqslant \alpha \norm{\varphi_j}$.
Now replace the $j$'th term in the decomposition of $\Theta ( x )$ by
the two terms $\ell_j ( x ) \Phi_{j , +} + \ell_j ( x ) \Phi_{j , -}$.
This yields a new decomposition of $\Theta$ with linearly independent
vectors and the upper bound is now multiplied by the factor
$2^{\frac{1}{p}} \alpha < 1$. Continuing this procedure, the bound on
the $p$-nuclearity index can be made arbitrarily small. Thus in
particular all finite rank operators would have vanishing
$p$-nuclearity index for $p > 1$. This argument would not be possible,
of course, if we also insisted that the vectors in $p$-nuclear
decompositions should belong to the range of the operator concerned,
and it is possible that this might yield a viable notion of
$p$-nuclearity.

On the other hand, the actual use that is made of the notion of
nuclearity in the literature, cf. \eg\ 
\cite{buchholz/wichmann:1986,buchholz/jacobi:1987,%
  buchholz/porrmann:1990,dantoni/hollands:2004}, hints at a slightly
different solution of the problems just indicated when Hilbert spaces
are considered. The calculation of upper bounds on the nuclearity
index is always performed by falling back on an orthonormal basis. So
a further possible attempt to overcome the difficulties for $p > 1$ is
to allow only nuclear decompositions in terms of an orthonormal basis.
In this case the 2-nuclearity index of an operator $\Theta : \Hscr_1
\rightarrow \Hscr_2$, $\Hscr_1$, $\Hscr_2$ Hilbert spaces, is just the
trace of the operator $\Theta \Theta^*$.
\begin{Pro}
  Let $\Hscr_1$ and $\Hscr_2$ be two Hilbert spaces and let $\Theta :
  \Hscr_1 \rightarrow \Hscr_2$ be a continuous operator. Then $\Theta$
  is a 2-nuclear operator in the sense of
  Definition~\ref{Def:p-nuclearity-phys} with only decompositions in
  terms of orthonormal bases allowed if and only if $\Theta \Theta^* :
  \Hscr_2 \rightarrow \Hscr_2$ belongs to the trace-class.
  The corresponding nuclearity index is given by
  \begin{equation}
    \label{eq:trace}
    {\norm{\Theta}_2}^2 = \trace ( \Theta \Theta^* ) \text{.}
  \end{equation}
\end{Pro}
\begin{proof}
  According to the definition there is an orthonormal basis
  $\set{\Phi_k}_{k \in \Nbb}$ such that
  \begin{align*}
    \Theta ( x ) = \sum_{k = 1}^\infty \scp{\Phi_k}{\Theta ( x )}
    \Phi_k & = \sum_{k = 1}^\infty \scp{\Theta^* \Phi_k}{x} \Phi_k
    \text{,} \quad x \in \Hscr_1 \text{,}\\
    \text{and} \quad \sum_{k = 1}^\infty \norm{\Theta^* \Phi_k}^2 & <
    \infty \text{.}
  \end{align*}
  But the expression on the left-hand side of the last inequality is
  just the trace of the operator $\Theta \Theta^*$, independent of the
  chosen orthonormal basis. On the other hand any operator $\Theta$
  with $\Theta \Theta^*$ lying in the trace-class allows for a
  2-nuclear decomposition in the sense of this Proposition. The
  equation~\eqref{eq:trace} is then an immediate consequence.
\end{proof}

The above discussion urges us to look for a more refined notion of
nuclearity to be applied to questions of physical interest. One
possible direction will be indicated in the next section, inspired by
our need for lower bounds on the nuclearity index in order to
establish quantum energy inequalities. It is worth noting that
Schumann \cite{schumann:1996} found applications of $( p > 1
)$-nuclearity (according to the mathematicians' definition) to the
physical question of statistical independence in quantum field theory,
and it would be of interest to reproduce these results with the
modified concept of nuclearity we envisage. However, one should be
aware that even in \cite{schumann:1996} the notions of $p$-nuclearity
and $p$-approximability are not always carefully distinguished. We
also note that Theorem 2.7 of \cite{schumann:1996} is trivial for $p >
2$, as the quantity estimated can be made arbitrarily small by a
variant of the dilution argument given above.

\section{\mathversion{bold}$\epsilon$-Entropy and Operator-Partition
  of Unity}

To proceed further in the direction towards establishing a close
relationship between the nuclearity condition and quantum energy
inequalities, one requires the following:
\begin{romanlist}
\item \label{list:independence} a satisfactory understanding at deeper
  levels of the natural reasons for the necessity to choose linearly
  independent vectors in $\Efrak$ and to pick up vectors belonging to
  the image set in $\Ffrak$ of the nuclear map $\Theta : \Efrak
  \rightarrow \Ffrak$, and,
\item \label{list:lower_bounds} sufficient control over lower bounds
  on an appropriate nuclearity index, as well as the upper bounds
  typically discussed in the literature.
\end{romanlist}

For this purpose, it appears promising to introduce such viewpoints as
\ref{list:entropy} the notion of $\epsilon$-entropy and
\ref{list:partition} the operator partition of unity in the general
context of \textit{rigged modules} over (possibly non-selfadjoint)
operator algebras formulated in the theory of operator spaces.
\begin{alphlist}
\item \label{list:entropy} According to \cite{akashi:1990}, the
  $\epsilon$-entropy of a compact positive operator $K : \Hscr
  \rightarrow \Hscr$ on a Hilbert space $\Hscr$ can be formulated as
  follows: because of the compactness, the image in $\Hscr$ of the
  unit ball $\Hscr_{1}$ under $K$ can be covered by a finite
  $\epsilon$-coverning: $K \Hscr_{1} \subseteq \bigcup_{i = 1}^{N} B(
  x_{i} ;\epsilon )$, whose minimum cardinality is denoted by $N( K ,
  \epsilon )$. Then, the $\epsilon$-entropy $S( K , \epsilon )$ of the
  compact positive operator $K$ is defined by
  \begin{equation}
    S( K , \epsilon ) \doteq \log_{2} ( N ( K , \epsilon ) ) \text{.}
  \end{equation}
  The upper and lower growth orders, $D ( K )$ and $d ( K )$, are
  defined, respectively, by
  \begin{equation}
    D ( K ) \doteq \underset{\epsilon
      \searrow0}{\overline{\lim}}\frac{\log S ( K , \epsilon
      )}{\log(1/\epsilon)};\text{ \ \ \ } d ( K ) \doteq \underset
    {\overline{\epsilon\searrow0}}{\lim}\frac{\log S ( K , \epsilon)}
    {\log(1/\epsilon)} \text{,}
  \end{equation}
  which means that $S ( K , \epsilon )$ is asymptotically bounded from
  above and below, respectively, by $e^{D ( K ) \log( 1/\epsilon )}$
  and $e^{d ( K )\log ( 1/\epsilon )}$:
  \begin{align}
    &  e^{d ( K )\log( 1/\epsilon
      )}\underset{\epsilon\searrow0}{\lesssim} S ( K , \epsilon
    )\underset{\epsilon\searrow0}{\lesssim}e^{D ( K ) \log ( 1/\epsilon
      )};\\ 
    &  d ( K )\log( 1/\epsilon
    )\underset{\epsilon\searrow0}{\lesssim}\log S ( K , \epsilon
    )\underset{\epsilon\searrow0}{\lesssim}D ( K )\log 
    ( 1/\epsilon ) \text{.}
  \end{align}
  The results presented in \cite{akashi:1990} are as follows:
  \begin{align}
    D ( K ) &  \doteq
    \underset{\epsilon\searrow0}{\overline{\lim}}\frac{\log 
      S ( K , \epsilon )}{\log( 1/\epsilon
      )}=\underset{n\rightarrow\infty}{\overline{\lim}}\frac{\log
      n}{\log( 1/\lambda_{n} )}\nonumber\\ 
    &  =\inf\set{p > 0 : \sum_{n = 1}^{\infty} \lambda_{n}^{p}=\trace
      ( K^{p} ) < \infty} \nonumber\\ 
    &  = \inf\set{p > 0 : K \in L^{p} ( \BH )} \text{;}\\
    d ( K ) & \doteq
    \underset{\overline{\epsilon\searrow0}}{\lim}\frac{\log 
      S( K , \epsilon )}{\log ( 1/\epsilon )}=\underset{\overline{\epsilon
        \searrow0}}{\lim}\frac{\log m ( K , \epsilon )}{\log ( 1/\epsilon )}
    = \underset{\overline{n \rightarrow \infty}}{\lim}\frac{\log n}{\log
      ( 1/\lambda_{n} )} \text{,}
  \end{align}
  where $m ( K , \epsilon ) \doteq \max \set{ n : \lambda_{n} >
    \epsilon}$ with $\lambda_{n}$ being the $n$'th largest eigenvalue
  of the compact operator $K$:
  \begin{equation}
    \label{eq:cpt} 
    K = \sum_{i} \lambda_{i} \ket{\xi_{i}} \bra{\xi_{i}} \text{.}
  \end{equation}
\item \label{list:partition} While the validity of the above Schatten
  decomposition \eqref{eq:cpt} looks to be restricted to the operators
  in a Hilbert space only, its essence can be carried over through the
  standard algebraic method of \textquotedblleft changes of
  rings\textquotedblright\ to far more general contexts of
  \textit{rigged modules} appearing in the theory of operator spaces,
  which can be summarised briefly as follows. First, the Banach
  spaces, $\Efrak$ and $\Ffrak$, respectively as the domain and target
  spaces of $\Theta$, can be replaced more appropriately in our
  context by the \textit{operator spaces}\footnote{After this paper
    was completed, F. Fidaleo kindly drew our attention to his article
    \cite{fidaleo:1994} which seems to be the first appearance of
    operator space techniques in algebraic QFT. Fidaleo uses
    non-commutative $L^{p}$ -spaces to reformulate the split property
    as a natural consequence of their operator space structures. Our
    aim here, however, is rather different; we seek an approach to the
    goals \ref{list:independence} and \ref{list:lower_bounds} above by
    means of an extension of $\epsilon$-entropy to the context of
    generalised compact operators on the rigged modules.} as
  \textquotedblleft quantised Banach spaces\textquotedblright\ whose
  concrete form can be understood as subspaces of operator algebras,
  $\Efrak \subseteq \BHone$ and $\Ffrak \subseteq \BHtwo$, and whose
  intrinsic characterisation is given in terms of the topology
  describing the \textit{complete boundedness} in terms of the norm
  $\norm{x}_{cb} \doteq \sup_{n} \norm{x}_{n}$ which should satisfy
  the following conditions \cite{effros/ruan:2000,pisier:2003}:
  \begin{align}
    \norm{\alpha \cdot x \cdot \beta}_{n} & \leqslant \norm{\alpha}
    \norm{x}_{n} \norm{\beta} \text{ \ \ \ (for } \forall \alpha , \beta
    \in M_{n} ( \Cbb ) , x \in M_{n} ( \Efrak ) ) , \\
    \bgnorm{ \left(
      \begin{array}
        [c]{cc}%
        x & 0\\
        0 & y
      \end{array}
    \right)}_{n+m} & \leqslant \max ( \norm{x}_{n} , \norm{y}_{m})
    \text{ \ \ \ (for }\forall x \in M_{n} ( \Efrak ) , \forall y \in
    M_{m} ( \Efrak ) ) \text{.}
  \end{align}
  A linear map $\Theta : \Efrak \rightarrow \Ffrak$ from an operator
  space $\Efrak$ to another $\Ffrak$ is completely bounded if it is
  bounded with respect to the spatial tensor norm,
  \begin{equation}
    \norm{\Theta}_{cb} \doteq \sup_{n \in \Nbb} \norm{ \Theta \otimes
    id_{M_{n}}} < \infty \text{,}
  \end{equation}
  where $\Theta \otimes \id_{M_{n}} : \Efrak \otimes M_{n} ( \Cbb ) =
  M_{n} ( \Efrak ) \rightarrow M_{n} ( \Ffrak )$. The total set of
  completely bounded operators from $\Efrak$ to $\Ffrak$ is denoted by
  $CB ( \Efrak , \Ffrak )$. To adapt these definitions to our present
  context, it is more convenient to restrict operator spaces to the
  \textit{rigged modules} defined as follows:
  \begin{Def}[Blecher \cite{blecher:1996}]
    A right $A$-operator module $Y$ with a (non-selfadjoint) operator
    algebra $A$ is called a \textit{\bfseries right $\mathib{A}$-rigged
      module} if there is a net of positive integers $n ( \beta )$ and
    right $A$-module maps $\phi_{\beta} : Y \rightarrow C_{n ( \beta
      )} ( A ) = A \otimes M_{n ( \beta ) , 1} ( \Cbb )$ and
    $\psi_{\beta} : C_{n ( \beta )} ( A ) \rightarrow Y$ such that
    \begin{arabiclist}
    \item $\phi_{\beta}$ and $\psi_{\beta}$ are completely contractive
      (\ie, contractive with respect to $\norm{~.~}_{cb}$);
    \item $\psi_{\beta} \phi_{\beta} \rightarrow \id_{Y}$ strongly on
      $Y$;
    \item the maps $\psi_{\beta}$ are right $A$-essential (\ie,
      $\overline{\linhull ( Y \cdot A )} = Y$);
    \item $\phi_{\gamma} \psi_{\beta} \phi_{\beta} \rightarrow
      \phi_{\gamma}$ uniformly in norm for $\forall \gamma$.
    \end{arabiclist}
  \end{Def}
  
  What is remarkable about this notion is not only that a Hilbert
  $C^*$-module as an operator-module analog of a Hilbert space is a
  rigged module in this sense, but also that an arbitrary rigged
  module can be embedded into a (possibly nonunique) Hilbert
  $C^*$-module as stated in the following theorem:
  \begin{The}[Blecher \cite{blecher:1996}]
    Let $A$ be a (non-selfadjoint) operator algebra with c.a.i. (=
    con\-trac\-tive approximate identity), $B$ a $C^*$-algebra
    generated by the unitisation $A_{+}$ of $A$ and $Z$ a Hilbert
    $C^*$-module over $B$. Suppose that $Y$ is a closed $A$-submodule
    of $Z$ and that $W \doteq \set{z \in Z : \scp{z}{y} \in A$ for
      $\forall y \in Y}$.  Suppose further that there exists a c.a.i.
    for $\Kbb ( Z )$ consisting of elements $\sum_{k=1}^{n}
    \ket{y_{k}} \bra{w_{k}}$ in $D \doteq \linhull \set{\ket{y}\bra{w}
      : w \in W \text{, } y \in Y }$ with $\sum_{k = 1}^{n}
    \ket{y_{k}}\bra{y_{k}} \leqslant 1$ and $\sum_{k = 1}^{n}
    \ket{w_{k}}\bra{w_{k}} \leqslant 1$. Let $C$ be the closure of $D$
    in $\Kbb ( Z )$. Then $Y$ is a right $A$-rigged module and $W$ is
    a left $A$-rigged module.  Moreover, $\tilde{Y} \cong \set{\bar
      {w} \in \bar{Z} : w \in W}$ as $A$-operator modules (completely
    isometrically)\ and $\Kbb ( Y ) \cong C$ completely isometrically
    isomorphically.  \textit{\bfseries Conversely, every rigged module
      arises in this way}.
  \end{The}
  In the above, the algebra $\Kbb ( Y )$ of generalised compact
  operators is defined by the norm limits in $\norm{~.~}_{cb}$ of
  finite-rank operators given by the linear combinations of $\ket{y}
  \bra{f}$, $y \in Y$, $f \in \tilde{Y} \doteq \bset{f \in CB ( Y , A )
  : f \text{ is $A$-linear and } (\psi_{\beta} \phi_{\beta})^{\ast} f
  \rightarrow f \text{ uniformly}}$, where $\tilde{Y}$ is the
  complete-boun\-dedness analog of the dual of $Y$. In terms of the
  Haagerup module tensor product $ \otimes_{h A}$
  \cite{effros/ruan:2000,pisier:2003} (as the most appropriate
  definition of tensor products in the contexts of operator spaces and
  of Hilbert modules), the natural relation $\Kbb ( Y ) \cong Y
  \otimes_{hA} \tilde{Y}$ holds. In this context our nuclear map
  $\Theta$ should belong to $\Kbb ( \Efrak , \Ffrak )$ which is also a
  rigged module and which can be embedded into the \textit{linking
    algebra }$\Kbb ( \Efrak \oplus \Ffrak )$ as its \textquotedblleft
  corner\textquotedblright\ entity:
  \begin{equation}
    \Kbb ( \Efrak , \Ffrak ) = \Ffrak \otimes_{hA} \tilde{\Efrak}
    \hookrightarrow \Kbb ( \Efrak \oplus \Ffrak ) = \left(
      \begin{array}
        [c]{cc}%
        \Kbb ( \Efrak ) & \Kbb ( \Ffrak , \Efrak )\\
        \Kbb ( \Efrak , \Ffrak ) & \Kbb ( \Ffrak )
      \end{array}
    \right)  \text{.}
  \end{equation}
  While the generalised compact operators are not compact operators in
  the genuine sense, they share many important features of the latter
  allowing the finite-dimensional approximations, which constitutes
  the essential ingredients of the `rig\-ged\-ness'. In this way, the
  essence of the Schatten decomposition $K = \sum_{i} \lambda_{i}
  \ket{\xi_{i}} \bra{\xi_{i}}$ can be recovered and generalised in the
  form of operator partition of unity: $\Theta = \sum_{i} \lambda_{i}
  \ket{x_{i}} \bra{y_{i}}$ on the basis of which a variety of
  \textit{entropy-like quantities} can be defined and calculated as
  already indicated by the above discussion of $\epsilon$-entropy (cf.
  Alicki's formulation of non-commutative dynamical entropy). Then,
  the essence of the quantum energy inequalities might perhaps be
  formulated as the stability condition imposed on the vacuum-like
  states in relation to the Legendre transform involving the energy
  (density) and one of the suitable entropy-like quantities (e.g.,
  $\alpha$-divergence and relative entropy \cite{amari/nagaoka:2001})
  which essentially originate from the type-III property of local
  subalgebras appearing in algebraic QFT.
\end{alphlist}

\paragraph{\itshape Acknowledgements}  I.O. and C.J.F. express their
sincere thanks to all the members of II. Institut f\"{u}r Theoretische
Physik at Universit\"{a}t Hamburg for their kind hospitality during
visits made in Summer and Autumn 2004, respectively, and in particular
to Dr B.  Kuckert for arranging financial support for these visits
under Emmy-Noether-Projekt Ku 1450/1--3. I.O. was also partially
supported by JSPS Grants-in-Aid (\#15540117). M.P. was funded by the
Deutsche Forschungsgemeinschaft (DFG), Emmy-Noether-Projekt Ku
1450/1--3, and wishes to thank the RIMS, Kyoto University, for
financial support and its staff for their kind hospitality in February
and March 2004. The reference to the paper of R. Schumann
\cite{schumann:1996} was kindly supplied by Professor D. Buchholz and
is herewith gratefully acknowledged. C.J.F.  thanks Dr S.P. Eveson for
a useful discussion relating to the various candidate $p$-nuclearity
conditions for $p > 1$.

\providecommand{\SortNoop}[1]{}

\end{document}